\documentclass[twocolumn,showpacs,preprintnumbers]{revtex4}%
\usepackage{graphicx}
\usepackage{dcolumn}
\usepackage{bm}
\usepackage{amsmath}
\usepackage{amsfonts}
\usepackage{amssymb}%
\begin{document}
\title{Quantum gravitational proton decay at high temperature}
\author{Ulf H. Danielsson}
\email{ulf@teorfys.uu.se}
\affiliation{Institutionen f\"{o}r teoretisk fysik, Uppsala Universitet, Box 803, 751 08
Uppsala, Sweden.}
\date{December 29, 2005}

\begin{abstract}
One of the most important challenges of contemporary physics is to find
experimental signatures of quantum gravity. It is expected that quantum
gravitational effects lead to proton decay but on time scales way beyond what
is of any relevance to experiments. At non-zero temperatures there are reasons
to believe that the situation is much more favourable. We will argue that at
the temperatures and densities reached at present and future fusion facilities
there is a realistic possibility that proton decay could be detectable.

\end{abstract}

\maketitle




\section{Introduction}

One way to violate the conservation of baryon number is to form a black hole
and let it evaporate through the emission of Hawking radiation \cite{H1}. The
formation and subsequent evaporation of a black hole does conserve charges
that are protected by local symmetries such as the electric charge. Baryon
number, on the other hand, is not conserved. Whereas the creation of a real
black hole is not possible in the laboratory, it has been argued that virtual
Planck size black holes are produced by quantum effects. This is the
space-time foam proposed by Wheeler \cite{W} and Hawking \cite{H2}. An
intriguing suggestion is that such virtual black holes can contribute to
baryon number violation and consequently to proton decay. In \cite{Z,HPP,AKMP}
the magnitude of such an effect was estimated, and the lifetime of a proton
was found to be given by roughly
\begin{equation}
\tau\sim m_{p}^{-1}\left(  \frac{m_{pl}}{m_{p}}\right)  ^{4}\sim
10^{45}\text{years},\label{lt}%
\end{equation}
where $m_{p}$ is the proton mass, and $m_{pl}$ is the Planck mass. This result
is many orders of magnitude above the estimates of theories of grand
unification and even further away from the experimental limits.

In this paper we will consider the possibility that the the life time of a
proton is substantially reduced at finite temperature. Our conclusion is that
the life time is of the order%
\begin{equation}
\tau\sim\frac{m_{pl}^{2}}{T^{3}}. \label{tid}%
\end{equation}
In section 2 we give two independent arguments for this result, one of which
is based on black hole complimentarity. If the temperature is high enough, our
formula predicts that the life time of a proton will be dominated by quantum
gravity. In section 3 we will argue that the conditions at present and planned
experimental fusion facilties are such that proton decay could be
experimentally detectable.

\section{Baryon number violation at finite temperature}

We will now argue for the existence of temperature dependent baryon number
violating processes due to quantum gravity and possibly related to the
production of virtual black holes. Even though the details of such processes
will have to wait for a fully developed theory of quantum gravity, it is
nevertheless simple to estimate their importance. We will present two quite
different arguments both yielding results in line with what is to be expected
on the grounds of simple dimensional analysis.

The key ingredient in our argument will be the assumed existence of an
effective smallest length of the order of the Planck length. We expect that
any distance shorter than this can not be given a meaning, and as a
consequence it is reasonable to expect that one can at most assign of the
order of one bit of information to the corresponding Planck volume. The state
of matter is encoded in the state of these bits, and in particular the state
of a proton will be described in this way. In a suitable basis it should be
possible to encode the baryon number of the configuration in one such bit in
particular. The way this is done might include non-local and holographic
aspects, details of which will not concern us. The value of this bit, and
hence the baryon number, would be expected to change in response to
interactions with an external agent such as a heat bath. By dimensional
analysis the characteristic interaction cross section can be estimated to be
of order $\sigma\sim l_{pl}^{2}$. In the presence of a heat bath there will be
a flux of, e.g., photons proportional to $T^{3}$, and as a consequence, we
estimate the interaction rate to be given by $\Gamma\sim l_{pl}^{2}T^{3}$ and
the life time is given by (\ref{tid}). A similar argument has been made in
\cite{DO} in the context of de Sitter space.

The argument is very simple but quite generic and it is reasonable to assume
that it holds true regardless of the details of quantum gravity. The only
thing we need is the presence of baryon number violation and that the rate is
naturally given in terms of the temperature and the strength of gravity.
Clearly, there will be other mechanisms contributing to baryon number
violation which sometimes will dominate. If the temperature is too small, for
instance, the decay rate will be governed by GUT-theory mechanisms. What we
argue is simply that (\ref{tid}) sets a universal upper limit on the life time
independent of the details of particle physics. As we will see below this
limit is on the one hand not in conflict with any experimental results, and on
the other hand within reach of realistic new experiments.

A different argument, providing further insights into the relevant physics,
can be obtained by investigating space times with horizons in more detail. To
this end we consider a probe in free fall towards the horizon of a black hole.
As is well known, a distant observer will not actually see the probe cross the
horizon but rather observe how the probe asymptotically approaches the
horizon, and how it becomes ever more redshifted and effectively invisible
after some time. If quantum effects are taken into account, the situation will
be slightly different since there in addition will be Hawking radiation
emitted by the horizon, \cite{H1}. While Hawking's original suggestion was
that this radiation did not carry any information about what ever had crossed
the horizon, \cite{H3}, the present generally accepted point of view is
different. It is believed that the radiation does carry all information with
it, albeit in a way difficult to decode. In fact, what a careful observer
would see when the probe approaches the horizon is how the probe is slowly
fried by the Hawking radiation. In this way the probe itself is dissolved into
radiation that carries all information about the probe. Note that we, in
principle, can follow what happens to the probe all along -- it will never
disappear from sight. For us it will be important to note that this dramatic
process does not conserve baryon number.

Confusingly, a second observer moving along with the probe will have a very
different version of what happened. The region interpreted as a horizon by the
first observer is nothing special to the second observer, who will just
continue the journey along with the probe. The resolution of the puzzle is by
now well known in the form of the black hole complementarity principle
\cite{STU}. The point is that the two observers will never be able to meet and
disagree about what happened precisely due to the existence of the horizon.

Now, how long does the process take? The horizon of a black hole has a
temperature of the order $1/R$ and an area given by $4\pi R^{2}$, where $R$ is
the radius of the black hole. (Our units are such that $h=c=1$.) The total
number of, e.g., photons emitted by the horizon per time unit is therefore
proportional to $1/R$. Most of these photons have nothing to do with the probe
we are tracking. Instead they are providing us with a picture of the complete
horizon, whose information content is given by the area of the horizon in
Planck units. Therefore we can not expect to clearly discern what has happened
to the probe until the number of received photons approaches this number. This
happens after a time of order $\tau\sim m_{pl}^{2} R^{3}$, which, we argue,
will be the time scale on which baryon number violation can be said to occur.
It should be noted that this coincides with what was argued in \cite{P} to be
the characteristic time before an evaporating black hole starts to deliver
information about its content.

From the point of view of the distant observer, the horizon of the black hole
appears as a surface heated to a temperature of $T\sim1/R$ where baryon number
violation is taking place. As far as the observer is concerned, the
unobservable region beyond the horizon is of no relevance. Given this
observation we conjecture, and this is a crucial step, that baryon number
violation is a consequence of the temperature of the horizon, and occurs
regardless of the source of the heat bath. Therefore the temperature need not
necessarily be associated with the Hawking radiation of a black hole, and we
propose that the life time of a proton in a heat bath at temperature $T$ is
given by (\ref{tid}).

Instead of focusing on the horizon of a black hole, one can construct a
similar argument using the cosmological horizon of de Sitter space. In de
Sitter space, the universe expands exponentially with time, and the distance
between two objects in free fall increases according to $e^{Ht}$, where $H$ is
a time independent Hubble constant. It is easy to show that any object further
away from a given observer than $1/H$ will forever remain invisible. This
distance will therefore represent a horizon analogous to the horizon of a
black hole. Again we can consider a probe in free fall towards the horizon and
again there will be radiation due to quantum effects \cite{GH}. The argument
proceeds in the same way as before, \cite{DO}, and again one finds a time
scale given by equation (2).

We have reached our estimate for the life time of a proton (or any other
particle with a conserved global quantum number) in two different ways.
Furthermore -- if there actually is a violation of baryon number due to
quantum gravity at finite temperature -- the expression we have arrived at is
more or less the simplest possible. In the next section we will investigate
the experimental consequences.

\section{Experimental predictions}

Given the independent arguments in the previous section, we are confident that
our estimate for the proton life time is correct. Though the result is
theoretically interesting, it would be even more important if the effects
could be experimentally detectable. In other words, can finite temperature
effects be important in present Earth based experiments looking for proton
decay? Using our formula and using a temperature of 300K one finds an
estimated proton life time on the order of $10^{40}$ years. This is shorter
than what is expected from effects already present at zero temperature, but
still several orders of magnitudes beyond the experimental limits.

What about reducing the proton life time by considering situations in which
the temperature is higher? An interesting example of high temperatures
sustained at a comparably long time is in the case of fusion reactors. In
present fusion research using, e.g., tokamaks, the criterion used to judge the
efficiency of the experiment is the Lawson criterion. The Lawson criterion
states that useful fusion will occur if $n\tau_{e}T\geq10^{21}$ $skeV/m^{3}$ ,
where $n$ is the number density of nuclei, $\tau_{e}$ is the containment time
of the plasma (i.e. the characteristic time it takes for the energy in the
plasma to dissipate), and $T$ is the necessary temperature, typically a few
times $10^{8}K\sim10keV$. At the present fusion experiment JET, one has almost
achieved this goal, \cite{JET}. Hence, with a volume for the plasma of around
$100m^{3}$, we can estimate that $N\tau_{e}\sim10^{22}s\sim10^{15}$years,
where $N$ is the total number of nuclei in the plasma. At the future fusion
reactor ITER, \cite{ITER}, one will satisfy the Lawson criterion, and with a
volume for the plasma of around $1000m^{3}$ , one gets $N\tau_{e}\sim
10^{23}s\sim10^{16}$years. These experimental parameters should be compared
with the predicted proton life time, which, using our argument from quantum
gravity, is given by $10^{20}$years at the temperatures relevant for fusion
reactors. We conclude that in case of JET one can expect on the order of one
event for every accumulated $10^{5}\tau_{e}$ of running time, whereas for ITER
the expected rate is one event for every accumulated $10^{4}\tau_{e}$ of
running time. A typical value for $\tau_{e}$ is one second.

The signature of proton decay is likely to be in the form of high energy
particles like pions, leptons, etc, with a characteristic total energy given
by the proton mass of $1GeV$ and a total unit positive charge. Our prediction
is that such events will take place at JET and ITER at the above calculated rates.

It should be noted that our estimate of the proton life time is very
approximate. For instance, we have used the Planck mass given by $m_{pl}%
\sim10^{19}GeV$, if instead we had used the reduced Planck mass, $m_{pl}%
/\sqrt{8\pi}\sim10^{18}GeV$, which is often used to estimate the importance of
quantum gravitational effects, we would have arrived at $10^{18}$ years. The
corresponding estimates for the necessary running times would have been two
orders of magnitude shorter.

While the detailed mechanism leading to the proton decay remains to be
explored, the argument presented in this letter is rather general. We
therefore propose that the necessary steps are taken to look for the effect at
present and future fusion facilities. A positive result would be of great importance.

\begin{acknowledgments}
The author is a Royal Swedish Academy of Sciences Research Fellow supported by
a grant from the Knut and Alice Wallenberg Foundation. The work was also
supported by the Swedish Research Council (VR).
\end{acknowledgments}

\end{document}